\documentstyle[preprint,epsf,aps]{revtex}
\begin{document}
\draft

\title{
 Singularities in droplet pinching with vanishing viscosity
    }
\author{
Jens Eggers
 \\Universit\"at Gesamthochschule Essen, Fachbereich Physik, \\ 
  45117 Essen, Germany}
\maketitle
\begin{abstract}
A slender-jet model for the pinching of a liquid column
is considered in the limit of vanishing viscosity. We find the model to
develop a singularity in the gradients of the local radius and
the velocity at a finite thread radius, so it does not 
describe breakup. However, the observed steepening 
of the profile corresponds to experiments and simulations 
with fluids at low viscosity.
The singularity has similarity form, which we compute analytically. 
The result agrees well with numerical simulations of the model
equations.

\end{abstract}
\newpage
\section{Introduction}
Considerable attention has been devoted recently to the breakup 
of an axisymmetric column of fluid 
\cite{/E/,/SBN/,/K/,/E3/,/DHL/}. The breakup is 
driven by surface tension forces, which reduce the surface area by 
contracting the fluid thread until its radius goes to zero at 
a point. As a typical example, we show a drop of water falling 
from a faucet in Fig.\ref{fig2} \cite{/PSS/,/ED/}. 

Close to the point of breakup, the interface looks like a cone 
attached to a nearly flat interface. This is characteristic for
low viscosity fluids, where viscosity is important only in a 
small spatial region around the point of breakup. Details of the initial 
conditions or external forces like gravity are believed to 
have little impact on
the very localized behavior close to pinch-off. Indeed experiments 
with or without gravity and for a variety of nozzle diameters
show very similar shapes \cite{/E3/}. This is because
surface tension forces become very strong near pinching, and 
drive very small amounts of fluid. Thus the very rapid motion
close to breakup is separated dynamically from the motion on the scale of
the nozzle diameter both in space and time. A proper measure
of length and time are the {\it local} scales 
\begin{equation}\label{1.1}
\ell_{\nu} = (\nu^2\rho)/\gamma \;,\; t_{\nu}=(\nu^3\rho^2)/\gamma^2,
\end{equation}
which only depend on the properties of the fluid. Here $\nu$ is the
kinematic viscosity, $\gamma$ the surface tension, and $\rho$ is
the density of the fluid. In the case that the minimum radius $h_{min}$ of
the fluid neck is much smaller than $\ell_{\nu}$, a universal 
pinching solution has been observed \cite{/E/,/K/}. As the radius 
of the fluid neck goes to zero, surface tension, viscous, and inertial
forces are of the same order.

For a low viscosity fluid like water, however, $\ell_{\nu}$
is only 100 \AA, so this asymptotic solution is hardly of relevance
experimentally. Accordingly, it would be extremely desirable
to develop a similarity theory valid in the range 
$h_{min} \gg \ell_{\nu}$. Assuming that all flow features of a 
hypothetical Navier-Stokes solution are of the same order as 
$h_{min}$, it is tempting to look at solutions of the inviscid (Euler) 
equation for that regime. Unfortunately, the Euler equation
is known \cite{/GMG/} to exhibit spurious blow-up of the local 
vorticity, even starting from smooth initial data. This problem can
be avoided by only considering the subclass of solutions which
are irrotational, and are thus described by potential flow. 
Numerical simulations of inviscid, irrotational flow driven
by surface tension were used by a number of authors \cite{/ML/,/S94/}
to describe pinching. After the minimum radius has reached 
a value of about 1/20 of its initial value, all simulations 
show an {\it overturning} of the profile. This means that the 
neck radius $h(z)$ is no longer a single-valued function of 
the position along the axis $z$. Recently \cite{/DHL/} it was shown
that this overturning can be understood as the convergence 
onto a universal similarity solution of the inviscid, irrotational 
equations. Similar findings were reported by Chen and Steen 
\cite{/CS/} for the related problem
of a soap film which drives a flow in the surrounding air. 
Neglecting the inertia of the film, this corresponds to the 
motion of two fluids of equal density with surface tension between 
them.

The fundamental question thus remains whether viscosity can really
be neglected. A Navier-Stokes computation at low viscosities
\cite{/AM/} showed no signs of overturning, although the initial
neck radius was $4 \cdot 10^{4} \ell_{\nu}$. Experimentally,
\cite{/PSS/,/BE/} there is also no sign of overturning even 
in water, an event that would show up in the side view as 
a flat wall perpendicular to the axis. Clearly, other and more
detailed simulations of low-viscosity fluids are needed to 
confirm this, but there is reasonable concern that the 
assumption behind the inviscid-flow calculations are 
fundamentally flawed. 

In \cite{/BE/} we use a slender-jet model, originally developed 
in \cite{/ED/}, to investigate the possible effect of a 
very small amount of viscosity on the pinching of a liquid 
thread. The model \cite{/BFL/,/ED/} can be shown to capture
all the leading-order contributions to the jet dynamics 
below the viscous scale $\ell_{\nu}$. In addition, it was 
demonstrated in \cite{/ED/} that by including the full mean
curvature as a higher-order contribution to the pressure,
the model can reproduce experimental profiles even when 
$h_{min} \gg \ell_{\nu}$. As an example, in Fig.\ref{fig2}
a few profiles calculated from the slender jet model are
superimposed on the experimental picture \cite{/ED/}.
Thus although the model is formally not valid in the 
presence of large gradients, it is interesting to understand the limit 
of small viscosity in this case since it turns out 
to be analytically tractable.

In \cite{/BE/} we show numerically that the slender-jet model 
for small viscosity develops very sharp gradients both in the 
local radius and the velocity field. This causes viscous effects 
to become important in a very small region where gradients are 
large even though the minimum radius $h_{min}$ is still much 
larger than $\ell_{\nu}$. Clearly this is because the small-viscosity 
dynamics produces flow features which are much smaller than 
$h_{min}$. 

In the present paper we show that the inviscid equations indeed
have a singularity which leads to a blow-up of gradients in 
finite time. This has been done by explicitly constructing a 
similarity solution which solves the equations for $\nu = 0$. 
We extend the numerical code used previously 
in \cite{/BE/} to even smaller viscosities to show that in the limit 
$\nu\rightarrow 0$ of small viscosities the slender-jet 
equations always select
this inviscid singularity. The blow-up occurs while the minimum radius is
still finite and an analytical description of the local profiles 
is given. In particular the inviscid equations are not able 
to describe breakup, although the full curvature term is kept.

If one follows the above argument, inviscid scaling theories 
\cite{/DHL/,/KM/,/TK/} may fail even for arbitrarily small $\ell_{\nu}$.
On the other hand, our analysis is based on 
a simplified one-dimensional description of a liquid thread,
which represents a serious 
shortcoming compared to the three-dimensional axisymmetric representation
of \cite{/DHL/}. In particular, overturning cannot be described
within the model. To settle the question whether inviscid 
scaling is consistent, a very careful comparison with both
experiment \cite{/PSS/,/BE/} and full Navier-Stokes simulations \cite{/AM/}
is necessary. In particular more carefully resolved Navier-Stokes
computations are highly desirable. 

In the next section we introduce the one-dimensional slender-jet 
model which forms the basis of our analytical description 
of the inviscid singularity. We then
present conclusive numerical evidence for the existence of 
a inviscid singularity at a time $t_c$. Derivatives
of the surface profile and of the velocity diverge like a power law
as function of $t_c - t$. In the third section we present an 
analytical theory of the inviscid singularity. The resulting surface profiles
agree well with numerical simulations.

\section{Model and simulations}
The main assumption underlying the model of axisymmetric free-surface 
flow to be considered here is that the fluid motion is directed 
mostly in the axial direction. This allows to set up an asymptotic 
expansion \cite{/BFL/} in the thread radius, which at leading order 
gives equations for the radius $h(z,t)$ of the thread and for the 
velocity $v(z,t)$, which only depend on the axial coordinate
$z$. In what follows we will deal with the model introduced
in \cite{/ED/}, 
\begin{mathletters}
\label{2.1}
\begin{eqnarray}
&& \partial_t h = -vh_z - v_zh/2 , \label{2.1a} \\
&& \partial_tv = -vv_z - p_z + 
\frac{3\nu}{LU}\frac{(h^2v_z)_z}{h^2},\label{2.1b} \\
&& p = \frac{1}{h(1+h_z^2)^{1/2}} - 
    \frac{h_{zz}}{(1+h_z^2)^{3/2}} , \label{2.1c}
\end{eqnarray}
\end{mathletters}
where the index refers to differentiation with respect to the variable.
The fields $h(z,t)$ and $v(z,t)$ have been non-dimensionalized using
some fixed length scale $L$ of the problem (like the radius of a nozzle). 
The length $L$ can be combined with surface tension $\gamma$ 
and density $\rho$ to make up a time scale $T=(\rho L^3/\gamma)^{1/2}$
and a velocity scale $U=L/T$. Every quantity to follow will be 
non-dimensionalized using these units.

Equation (\ref{2.1a}) expresses mass conservation for a radially 
uniform velocity field. Conservation of momentum (\ref{2.1b}) not
surprisingly has the form of Burgers' equation in the inviscid limit,
driven by surface tension forces which are proportional to the mean 
curvature (cf. (\ref{2.1c})). Along the same lines of reasoning 
the equations (\ref{2.1}) were guessed by Lee \cite{/L/} for
$\nu = 0$. By including the full mean curvature in (\ref{2.1c}) 
we have gone beyond the leading order asymptotics to exactly 
reproduce the static shape of a hanging drop suspended from
an orifice. As an additional benefit, the most dangerous 
short-wavelength instabilities 
of the leading order model $p = 1/h$ have been removed. 
In fact, the leading order model is elliptic for $\nu=0$ 
\cite{/P95/,/FV/} and is thus ill-posed as an initial 
value problem. 

In \cite{/ED/,/BE/} a finite difference scheme was developed 
capable of simulating (\ref{2.1}) at very low viscosities. 
To resolve the small-scale structures we are interested in,
it is crucial to use an adaptive scheme, both in time and 
space. The minimum thread radius and the maximum gradient 
of $h$ were taken as predictors where additional spatial 
resolution was necessary. Thus in a typical run the grid spacing 
at the position of the inviscid singularity was 6 orders of 
magnitude smaller than at the boundary of the computational 
domain. Since the equations at low viscosity are very sensitive 
to noise, grids with smoothly varying grid spacings had to 
be used, where the spacing did not change by more than 1 \% 
from one grid point to the next. With these precautions, no 
numerical damping or dissipation had to be used, except for the 
physical viscosity $\nu$. 

Since we are interested in the limit of small viscosity, it 
would be tempting to put $\nu = 0$ in (\ref{2.1b}) directly. 
However, we found that as soon as the motion is sufficiently 
nonlinear, our scheme developed instabilities on the scale of 
the grid, which caused the code to break down. Thus the inclusion
of the full curvature term in (\ref{2.1c}) is not enough to
stabilize the numerical scheme. Similar short-wavelength
instabilities have also been reported in \cite{/S3/} using a
finite-element approach. On the other
hand, exceedingly small amounts of viscosity are sufficient 
to stabilize the scheme, even though the viscous term is smaller
than the others by several orders of magnitude throughout 
the domain. In the following, when speaking of a numerical solution of
the inviscid equations, we will always refer to the {\it limit}
of zero viscosity at {\it constant time}. 

We also experimented extensively with other means of regularizing 
the inviscid equations, for example by using numerical viscosities 
as in \cite{/ED/}. In the upwind differencing scheme introduced in \cite{/ED/},
the numerical viscosity is proportional to the grid 
spacing. The hope was to develop a scheme which automatically 
converges to the inviscid limit as one increases the resolution. 
Indeed, if the grid is coarse, a numerical viscosity was often sufficient 
to remove instabilities. But with improved resolution we always found 
the instabilities to return. Thus keeping a finite viscosity turns
out to be the only reliable and at the same time the most physical 
way of dealing with the instabilities. 
These results indicate that the system (\ref{2.1}) might be 
an ill-posed initial-value problem, in spite of the short-wavelength
reguralization introduced by (\ref{2.1c}). If on the other 
hand the problem is well-posed, and the instabilities are a 
problem of the numerical scheme, the above limit of small 
viscosity will yield a solution which coincides with the 
one defined by the inviscid (Lee's) equations. 

It is clear that the slender-jet approximation represented by
(\ref{2.1}) {\it assumes} the existence of a single-valued 
function $h(z)$ and is thus not able to describe overturning
as observed for three-dimensional axisymmetric, 
irrotational flow \cite{/ML/,/S94/,/DHL/}.
Thus there is no direct correspondence to the pinching solution found 
in \cite{/DHL/}. It could be that the inviscid singularity 
described in this paper is a consequence of the three-dimensional equations 
overturning. On the other hand, as explained in the Introduction, 
our working assumption is that the inviscid limit of the 
Navier-Stokes equation is singular,
in which case the limit of vanishing viscosity has no 
connection with solutions of the Euler equation. Thus the 
fact that full the potential flow equations show overturning
has no obvious implications for the limit of of vanishing 
viscosity studied here. The resolution of these issues must 
clearly await full Navier-Stokes computations at small viscosity.

Figure \ref{fig3a} shows a simulation of (\ref{2.1}) at a very
large Reynolds number in a liquid bridge geometry with initial 
radius $r=0.05$ and a small sinusoidal perturbation of
wavelength $\lambda = 4\pi r_0$ superimposed on it. Owing to the 
Rayleigh instability, the bridge starts to pinch. Shown are 
three profiles close to the inviscid singularity, where the minimum 
radius has already decreased by a factor of 20. To the right of
the minimum, an almost conical neck region is seen, on the other 
side a round drop has formed. 
Because of its small radius, the pressure 
in the flat region is high, pushing fluid over to the right. 
This causes the interface to perform a sliding motion, which lets
the interface become steeper and steeper, since the drop cannot move 
appreciably owing to its large inertia.
Note that the pressure goes to a value close
to zero at the eventual place 
of the inviscid singularity, marked by an arrow. 

Next we zoom in on the point around which $h_z$ goes to 
infinity, marked by an arrow in Fig.\ref{fig3a}. To sufficiently 
resolve the singularity without letting viscosity become 
important, the Reynolds number is chosen to be $Re = 4.7\cdot10^9$,
much larger than the Reynolds numbers treated in \cite{/BE/}.
With such a small Reynolds number we are no longer able to
resolve the huge range of length scales between the outer and
the viscous scale, because $L/\ell_{\nu}$ is now $2\cdot10^{19}$.
However, the early stages of the evolution of the liquid bridge,
where viscous effects are still small can safely be resolved. 
In \cite{/BE/} it was demonstrated that the slope goes to 
infinity near the inviscid singularity and thus the singularity 
time $t_c$ can be estimated from the blow-up of $h_z$. 
It follows from our scaling theory, to be presented in the 
next section, that $(h_z)_{max} \sim (t_c - t)^{-1}$. Thus 
$t_c$ can be computed very accurately by plotting 
$((h_z)_{max})^{-1}$ versus time and fitting with a linear law.
In Fig.\ref{fig4} we plot the maximum pressure gradient $p_z$,
which drives the fluid motion, and the maximum velocity 
gradient as a function of $t_c - t$. 
It is seen that both $p_z$ and $v_z$ settle on a power law 
\begin{equation}\label{2.5}
(p_z)_{max} \sim (t_c - t)^{-1} ,\quad (v_z)_{max} \sim (t_c - t)^{-1} .
\end{equation}
The pressure gradient contains the highest (third) derivative 
in the problem. The fluctuations seen in the curve thus give
an estimate of the amount of noise introduced by the regridding 
procedure. No noise is seen in the first velocity derivative, 
which clearly confirms the scaling given in (\ref{2.5}).
We thus see that the inviscid singularity is governed by power law scaling,
which will be investigated in more detail in the next section. 
We made sure that for all of the evolution shown the viscous 
term remained small compared to other terms. Thus the observed 
scaling represents the inviscid limit of the equations we want
to investigate. 

We also considered a great variety of other initial perturbations,
which changes the initial evolution of the bridge. In addition, 
we considered the bifurcation of a hanging drop in the presence
of gravity as shown in Fig. \ref{fig2}. In all these different 
cases, we always found the same singularity with
scaling (\ref{2.5}) to be selected in the inviscid limit. 
Thus there is good evidence that the inviscid singularity 
to be described below is a universal feature of the equations, 
independent of initial conditions. 

The equations (\ref{2.1}) with ($\nu = 0$) or 
systems very similar in structure have been used 
by a number of researchers \cite{/TK/,/L/,/M/,/MS/,/S2/,/S3/,/P95/,/PO/}
to describe pinching. However in Lee's original paper \cite{/L/} 
and most of the later work no attempt is made to resolve 
any detailed structure in the pinch region. 
For example, in units of the length of the computational 
domain, the grid spacing is $dx = 1/20$ in \cite{/L/} and 1/50 in \cite{/M/}.
For comparison, the minimum grid spacing used in the present 
paper is $dx_{min} = 10^{-10}$. In most papers
\cite{/L/,/M/,/MS/,/S3/} the computation is stopped 
at $h_{min}$ between 1/10 and 1/20 of the initial radius, when 
the position of the minimum and the drop size can be faithfully 
estimated. In \cite{/S3/}, which uses a finite element code,
the computations were stopped when numerical instabilities on the 
scale of the grid were observed. A spectral method is used in
\cite{/PO/}, but the computation was stopped at $h_{min} = 0.3$
due to numerical problems. Thus our finding of a short-wavelength
instability of the inviscid equations is consistent with other
numerical work, using a variety of different methods. It remains to 
be seen whether the inviscid versions of the Cosserat model
\cite{/M/} or other asymptotic models \cite{/PO/} are more 
well behaved. 

\section{Inviscid similarity solution}
We have seen in the previous section that derivatives grow sharply 
near the inviscid singularity, while on the other hand the height and
the velocity remains finite. This means that the similarity ansatz of 
\cite{/E/} has to be generalized to include a ``background'' height 
and velocity profile, which is slowly varying on the scale of the singular
part. At the same time, the singularity may be moving with some speed 
$V_s$, which is not necessarily the speed $V$ with which it is convected.
Thus one ends up with the similarity form
\newpage
\begin{eqnarray}\label{3.1} 
&& h(z',t') = H + t'^{\alpha}
f\left(\frac{z'+V_st'}{t'^{\beta}}\right) \nonumber \\
&& \\
&& v(z',t') = V + t'^{\alpha}
g\left(\frac{z'+V_st'}{t'^{\beta}}\right), \nonumber 
\end{eqnarray}
where 
\begin{equation}\label{3.2} 
z' = z - z_c \;\mbox{and}\; t' = t_c - t 
\end{equation}
measures the spatial and temporal distance from the singularity, 
respectively. On the spatial scale on which the inviscid 
singularity develops, $H$, $V$, and $V_s$ are approximately constant.
Note that (\ref{3.2}) has self-similar form, which is superimposed
on a traveling wave solution. 
Also, we assumed $h(z',t')$ and $v(z',t')$ to have the same 
scaling exponents, because this automatically balances the terms 
$h_z v$ and $v_z h$ in (\ref{2.1a}). 

For the ansatz (\ref{3.1}) to be consistent, one needs $\alpha > 0$, 
so in the singular limit $t' \rightarrow 0$ one is left with 
the finite height $H$ and velocity $V$. For the derivatives to blow
up, $\beta > \alpha$. To see 
whether (\ref{3.1}) solves the model equations (\ref{2.1}),
we balance the most singular terms in $t'$, deriving equations in the 
similarity variable 
\begin{equation}\label{3.3}
\eta = \frac{z' + V_st'}{t'^{\beta}} .
\end{equation}
One thus finds from (\ref{2.1a})
\begin{equation}\label{3.4}
(\beta f'\eta - \alpha f)t'^{\alpha-1} = 
-\left[(V-V_s)f' + \frac{1}{2} H g'\right] t'^{\alpha-\beta} .
\end{equation}
We expect the right hand side of (\ref{3.4}) to make the dominant 
contribution, which will be the case if $\beta > 1$. This is because 
then the function $f$ drops out of the equation, and the 
similarity equations only depend on the derivatives $f'$ and $g'$. 
Thus both $f$ and $g$ are determined only up to constants, which is
needed for consistency because our ansatz (\ref{3.1}) 
has a free constant built in.

Consequently, the angular bracket must vanish, giving
\begin{equation}\label{3.5}
g' = \frac{2(V_s-V)}{H} f' ,
\end{equation}
which means that up to constants and a difference in amplitude 
$2(V_s-V)/H$ the profiles of the height and of the velocity 
are the same.

Turning to (\ref{2.1b}) with $\nu = 0$, one finds to leading 
order 
\begin{equation}\label{3.7}
-(V_s-V)g't'^{\alpha-\beta} = \left(\frac{f''}{f'^3}\right)' t'^{-2\alpha} .
\end{equation}
Note that the term on the left corresponds to the highest derivatives 
in $p_z$ as given by (\ref{2.1c}). Balancing the left and the right 
hand side, one finds the scaling law
\begin{equation}\label{3.8}
\beta = 3\alpha .
\end{equation}
Combining (\ref{3.5}) and (\ref{3.7}), the similarity equation reads 
\begin{equation}\label{3.9}
-af' = \left(\frac{f''}{f'^3}\right)' ,
\end{equation}
where $a = 2(V_s - V)^2/H$. Evidently, the constant $a$ can be eliminated by 
the transformation 
\begin{equation}\label{3.10}
\phi(\eta) = a^{-1/3} f'(\eta).
\end{equation}
The most general solution of the equation for $\phi$,
\begin{equation}
\label{pequ}
-\phi = \left(\frac{\phi'}{\phi^3}\right)',
\end{equation}
has the form
\begin{equation}\label{3.11}
\phi(\eta) = \phi_0 F[\phi_0^{3/2}(\eta - \eta_0)] ,
\end{equation}
where $F(\xi)$ is a particular solution of (\ref{pequ}).

Equation (\ref{3.9}) can easily be solved using standard \cite{/BO/}
tricks. In view of the freedom implied by (\ref{3.11}), we choose 
$F(\xi)$ to have its maximum at $\xi=0$, and to fall off to $1/2$ 
at $\xi = \pm 1/2$. Then $F$ is given implicitly by
\begin{equation}\label{3.12}
\xi = \frac{1}{8F^{3/2}} (1 + 2F)(1 - F)^{1/2} .
\end{equation}
This function is represented in Fig. \ref{fig5} as the solid line. 
It decays to zero like $\xi^{-2/3}$ at infinity. In view of the similarity 
form (\ref{3.1}) this ensures that the leading dependence on $t'$ drops
out far away from the singular point. This is necessary for the solution 
to match onto the slowly varying background field.  

We note that the singularity described above is not just kinematic
in nature, since a contribution from the capillary forcing
enters the dominant balance in (\ref{3.7}). In addition, the 
form (\ref{3.1}) of the singularity with $V_s \neq V$ implies that
it is also a traveling wave. Surprisingly, the local shape 
(\ref{3.12}) of the first derivative of the local radius 
is identical to that of a scaling
solution $h = t'G(z'/t'^{3/2})$ of the simple kinematic wave
$\partial_t h + h\partial_z h = 0$ \cite{/R/}. This can be checked directly
from the similarity equation for $G(\xi)$. At present we do not know 
if this is a coincidence or is the result of a deeper analogy. 

To test the prediction of the theory, we took the same simulation 
as in Fig. \ref{fig4} at a time where the slope was $h_z = 10^{4}$.
We included the profile of $h_z$ in Fig. \ref{fig5}, shifting the 
maximum to the origin and normalizing its width. It is evident that
our analytical theory is fully confirmed by the comparison. 
The free constants $\phi_0$, $\eta_0$ in (\ref{3.11}) and the parameter 
$a$ in (\ref{3.9}) are not determined by the similarity theory. 
Indeed, we confirmed that they depend on initial conditions and 
therefore cannot come out of a local theory.  

Our next concern is to find the scaling exponents $\alpha$ and $\beta$, 
which are not determined from dimensional reasoning as in \cite{/E/},
but which are constraint by the scaling relation (\ref{3.8}). 
In addition, (\ref{3.9}) does not depend on the values of the exponents, 
so $\alpha$ cannot be selected by properties of the similarity equation,
as was the case in \cite{/UE/}.
To investigate this problem, we must look at next to leading order terms 
such as the ones contained in Eq. (\ref{3.4}). Correspondingly, 
there are sub-leading terms in $h$ and $v$, which have the form
\begin{equation}\label{3.13}
h(z',t') = H + t'^{\alpha}
f\left(\frac{z'+V_st'}{t'^{\beta}}\right) + 
t'^{2\alpha} f_1\left(\frac{z'+V_st'}{t'^{\beta}}\right) + \dots,
\end{equation}
and correspondingly for $v(z',t')$. Then (\ref{2.1a}) becomes
to next to leading order:
\begin{equation}\label{3.14}
\left[3\alpha f'\eta - \alpha f\right]t'^{\alpha-1} = 
\left[-gf'-\frac{1}{2}fg'+(V_s-V)f_1'-\frac{H}{2}g_1'\right]t'^{-\alpha}.
\end{equation}
Since the terms must balance, we get 
\begin{equation}\label{3.15}
\alpha=\frac{1}{2},\quad \beta = \frac{3}{2} ,
\end{equation}
which are the desired exponents. Note that this conforms with
the scaling of both the pressure and the velocity gradient 
from (\ref{2.5}), 
since
\begin{equation}
p_z \approx  t'^{-2\alpha}\left(\frac{f''}{f'^3}\right)' \quad\mbox{and}
\quad v_z \approx t'^{\alpha-\beta} g'.
\end{equation}
Thus both the value of the exponents and the shape of the 
profiles is in excellent agreement with theory. 

\section{Discussion}

In \cite{/BE/} we have studied the steepening of the height profile 
for small but finite viscosities. The slopes saturate at a 
large value, the maximum slope roughly following a scaling law
$(h_z)_{max} \sim Re^{1.25}$. As long as the viscous term is much 
smaller than the pressure gradient, one can use the inviscid 
similarity solution. The naive expectation is
that the slope saturates as soon as the viscous term is of the same 
order as the pressure gradient. Using (\ref{3.1}) and (\ref{3.15}),
the temporal scaling of the pressure gradient is $p_z \sim t'^{-1}$ and 
that of the viscous term $Re^{-1}(v_zh)_z/h^2 \sim Re^{-1} t'^{-5/2}$,
where $Re^{-1} = \nu/(UL)$ is constant.
Equating the two we find $(h_z)_{max} \sim Re^{2/3}$, which is
far too small an exponent. A possible explanation is that the Reynolds
numbers for which the exponent $1.25$ was found are still too small. 
But more likely there is an intricate interplay between 
the inviscid singularity and viscosity, leading to a more complicated 
intermediate scaling range. Indeed, for the Reynolds numbers considered, 
the slope continues to grow far beyond the time where the pressure 
gradient first balances the viscous term at a point. 
It thus remains a challenge to find the mechanism which makes 
the slope saturate. 

The pinching of vortex sheets and jets within the framework of 
inviscid hydrodynamics has recently been the object of much scrutiny
\cite{/CLS/,/PS/,/LT/}. 
The motion is either driven by surface tension, or 
comes from the flow field generated by the regions of high
vorticity. Short wavelength instabilities arise just as in 
our problem, which are usually 
removed by filtering out high-wavenumber components 
of the Fourier spectrum \cite{/Kr/}. It would be interesting to see 
whether this is equivalent to adding a small amount of viscosity, 
which is the physical regularization used in the present paper.
It is not clear that Krasny 
filtering actually selects the physical solution.
So far, analytical solutions have not been found for the 
singular motion of vortex sheets or lines. It remains to be seen
whether the singularity structure found in the present 
problem carries over.

We have seen that the slender-jet model (\ref{2.1}) in the limit of
small viscosity is characterized by by more than just the scale of
the minimum radius. Instead, a shock-type singularity develops 
whose width represents another, much smaller scale. The crucial 
question is of course whether a similar mechanism is at work 
in the small-viscosity limit of the Navier-Stokes equation, 
which could make this limit singular. 
It is unlikely that the three-dimensional equation has precisely
the same spatial singularity structure than the one found in the 
model equations, which constrain axial velocity gradients
to a far greater extend. Instead it is probable that high-pressure
fluid in the neck is injected into the drop, a situation which
is only poorly captured by the slice average of the model.
The Navier-Stokes equation might thus form a very thin boundary 
layer around the jet, in which viscosity remains important 
even for arbitrarily small $\nu$. This is a possible scenario
which would invalidate the purely inviscid calculation of
\cite{/DHL/}, at least form a physical point of view.

Even for the model equations, many 
unanswered questions remain. It is fascinating that even a simple 
one-dimensional model is capable of such complexity, 
the key to its understanding lying in the analysis of the 
singularities.

\acknowledgements
I have benefited greatly from discussions with Michael Brenner.
This work is partially supported by the Deutsche Forschungsgemeinschaft
through Sonderforschungsbereich 237.

\clearpage

\begin{figure}[h]
\begin{center}
\leavevmode
\epsfsize=0.5 \textwidth
\end{center}
\caption{\label{fig2} 
A drop of water falling
from a faucet $0.51cm$ in diameter. The lines 
represent a computation using one-dimensional model equations.
This figure is taken from \protect \cite{/ED/}.
       }
\end{figure}

\begin{figure}[h]
\begin{center}
\leavevmode
\epsfsize=0.6 \textwidth
\end{center}
\caption{\label{fig3a} 
A closeup view of simultaneous radius, velocity, and pressure 
profiles close to the inviscid singularity. The Reynolds number
is $Re = 5\cdot10^9$, and the time between successive profiles 
is $5\cdot10^{-5}$. The arrow marks the asymptotic location of the
shock.
   }
\end{figure}

\begin{figure}[h]
\begin{center}
\leavevmode
\epsfsize=0.9 \textwidth
\end{center}
\caption{\label{fig4} 
The maximum of the gradient of the pressure and of the velocity
as function of the 
time distance from the inviscid singularity. The Reynolds number is 
$4.7\cdot10^9$. Both curves asymptote to a slope very close to
$-1$.
   }
\end{figure}

\begin{figure}[h]
\begin{center}
\leavevmode
\epsfsize=0.9 \textwidth
\end{center}
\caption{\label{fig5} 
The normalized similarity function $F(\xi)$, cf. (\protect{\ref{3.10}}),
(\protect{\ref{3.11}}).
    }
\end{figure}


\newpage
\begin{center}
{\Huge Figure 1}
\end{center}
\begin{figure}[t]
\begin{center}
\leavevmode
\epsfsize=0.5 \textwidth
\epsffile{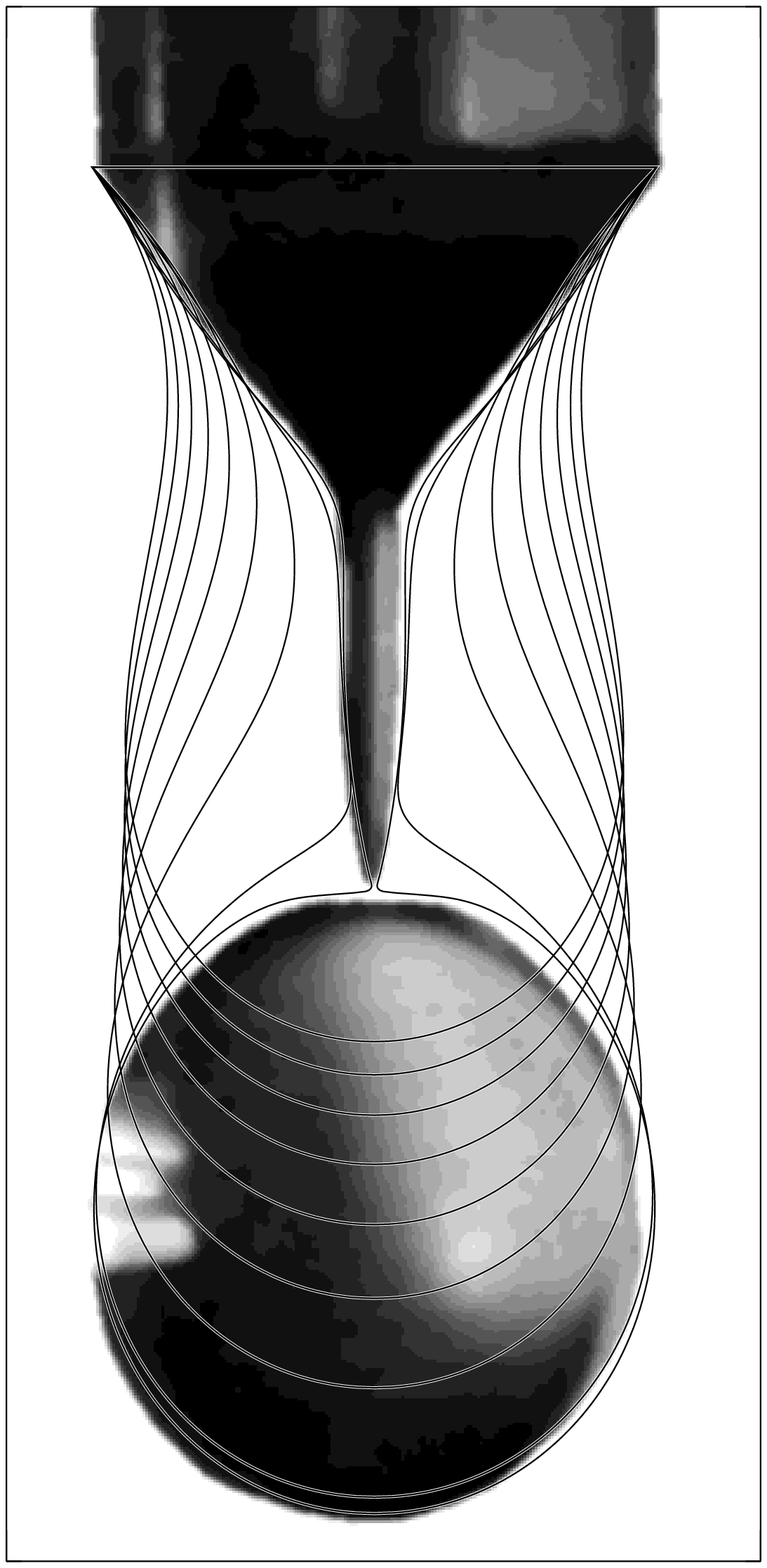}
\end{center}
\end{figure}

\newpage
\begin{figure}[h]
\begin{center}
  {\Huge Figure 2}
\vskip2cm
\leavevmode
\epsfsize=0.9 \textwidth
\epsffile{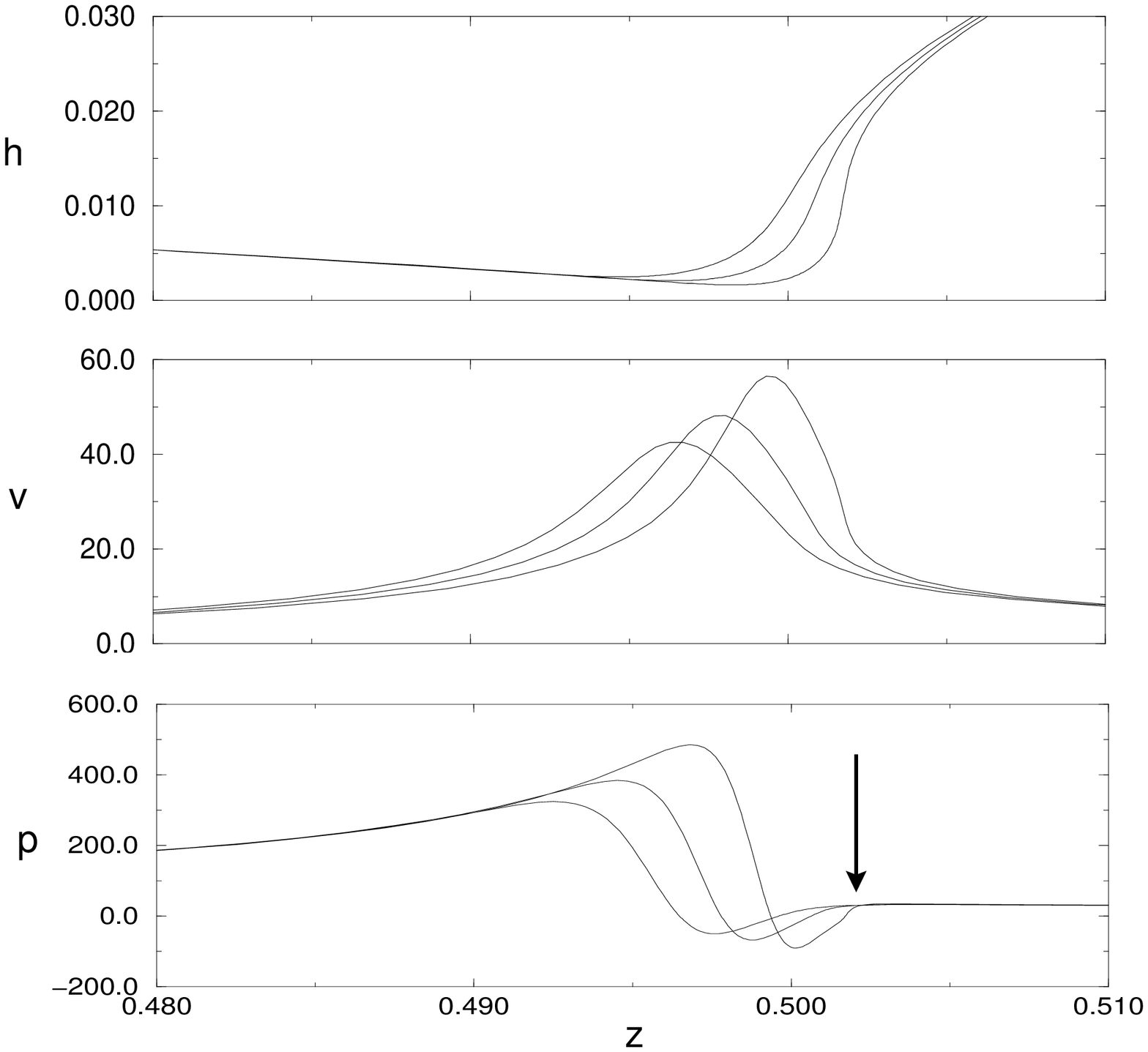}
\end{center}
\end{figure}

\newpage
\begin{figure}[h]
\begin{center}
  {\Huge Figure 3}
\leavevmode
\epsfsize=0.9 \textwidth
\epsffile{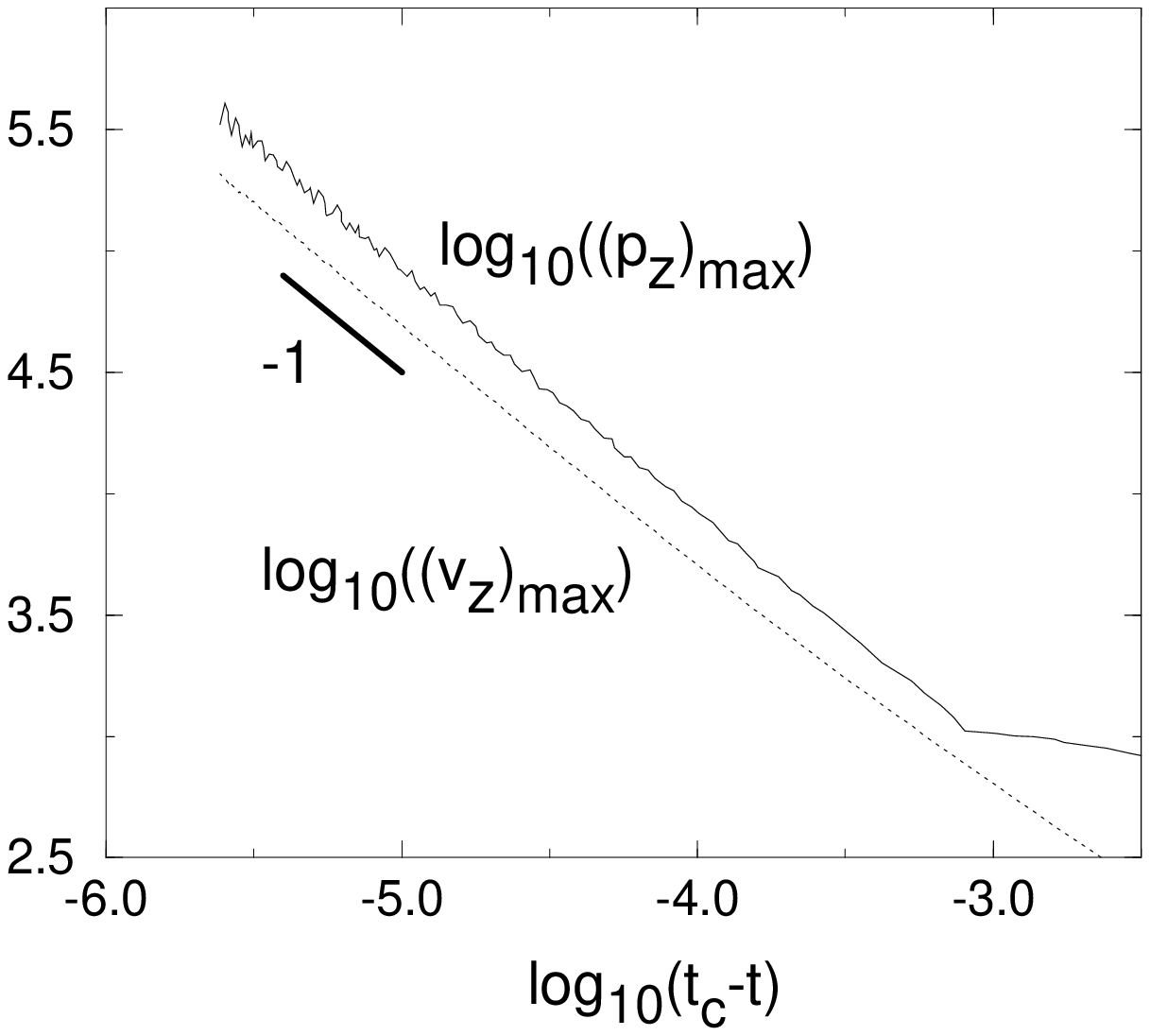}
\end{center}
\end{figure}

\newpage
\begin{figure}[h]
\begin{center}
  {\Huge Figure 4}
\leavevmode
\epsfsize=0.9 \textwidth
\epsffile{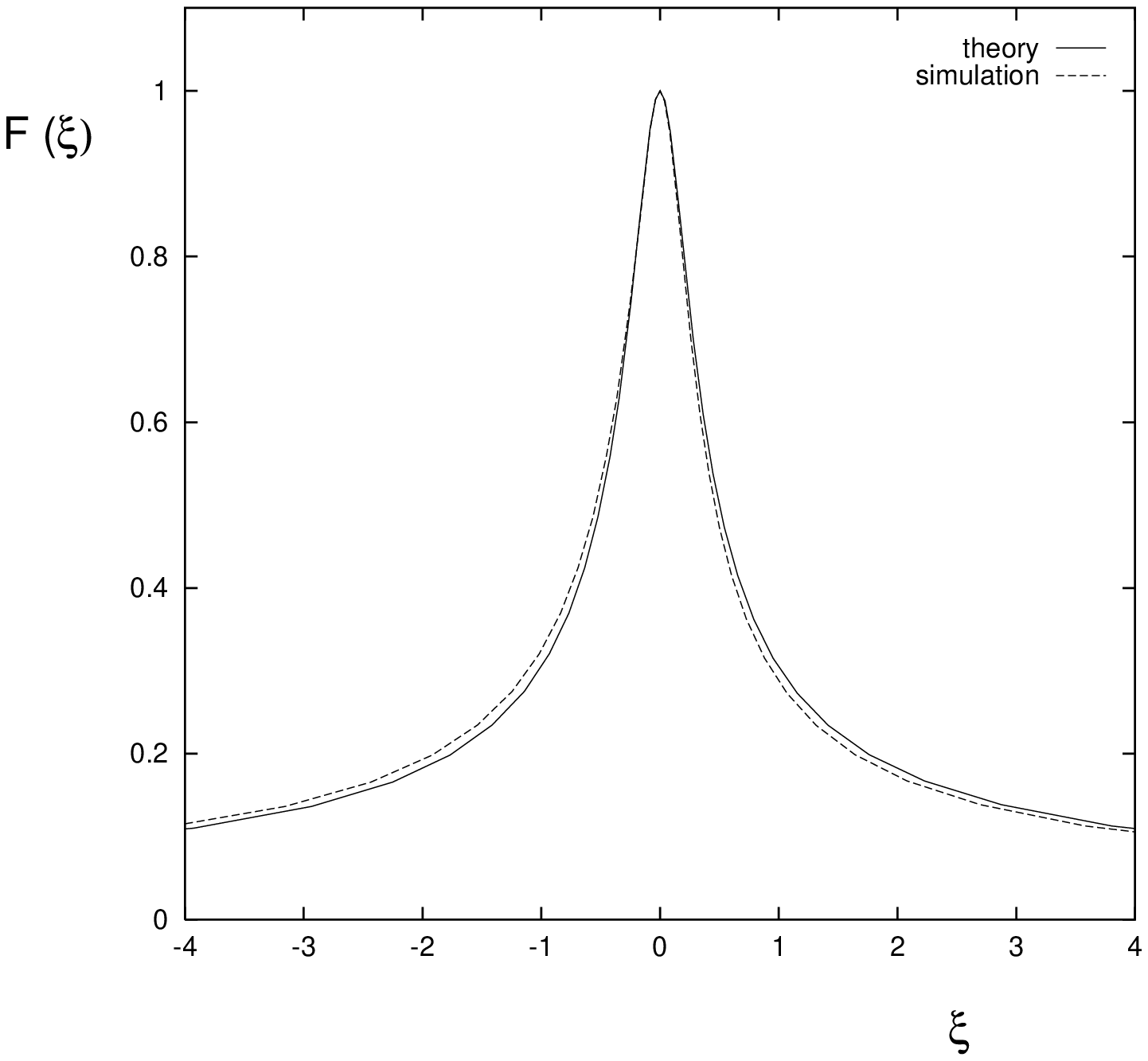}
\end{center}
\end{figure}


\begin{references}

\bibitem{/E/} J. Eggers, ``Universal Pinching of 3D Axisymmetric
Free-Surface Flow,'' Phys. Rev. Lett. {\bf 71}, 3458 (1993).  

\bibitem{/SBN/} X. D. Shi, M. P. Brenner, and S. R. Nagel,
``A Cascade Structure in a Drop Falling from a Faucet,''
Science {\bf 265}, 157, (1994)

\bibitem{/K/} T. A. Kowalewski, ``On the separation of droplets
from a liquid jet'', Fluid Dyn. Res. {\bf 17}, 121 (1996).

\bibitem{/E3/} J. Eggers, ``Nonlinear dynamics and breakup of free-surface
flows'', Rev. Mod. Phys. {\bf 69}, 865 (1997).

\bibitem{/DHL/} R. F. Day, E. J. Hinch, and J. R. Lister, 
``Self-similar Capillary Pinchoff of an Inviscid Fluid'',
Phys. Rev. Lett. {\bf 80}, 704 (1998).

\bibitem{/PSS/} D. H. Peregrine, G. Shoker, and A. Symon, ``The
bifurcation of liquid bridges,'' J. Fluid Mech. {\bf 212}, 25 (1990).

\bibitem{/ED/} J. Eggers and T. F. Dupont,
``Drop Formation in a One-Dimensional Approximation
of the Navier-Stokes Equation,'' J. Fluid Mech. {\bf 262}, 205 (1994).

\bibitem{/GMG/} R. Grauer, C. Marliani, and K. Germaschewski,
``Adaptive mesh refinement for singular solutions of the incompressible
Euler equations'',
Phys. Rev. Lett. {\bf 80}, 4177 (1998).

\bibitem{/ML/}
N.N. Mansour and T.~S. Lundgren,
``Satellite formation in capillary jet breakup'',
Phys. Fluids {\bf 2}, 114, (1990).

\bibitem{/S94/}
R.~M. S.~M. Schulkes,
``The evolution and bifurcation of a pendant drop'',
J. Fluid Mech. {\bf 278}, 83, (1994).

\bibitem{/CS/} Y.-J. Chen and P. H. Steen,  
``Dynamics of inviscid capillary breakup: collapse and pinch-off 
of a film bridge'',
J. Fluid Mech. {\bf 341}, 245 (1997).

\bibitem{/AM/}
N. Ashgriz and F. Mashayek, 
``Temporal analysis of capillary jet breakup'',
J. Fluid Mech. {\bf 291}, 163, (1995).

\bibitem{/BE/} M. P. Brenner et al., ``Breakdown of scaling in droplet 
fission at high Reynolds numbers'', Phys. Fluids {\bf 9}, 1573 (1997).

\bibitem{/KM/} J. B. Keller and M. J. Miksis, 
``Surface Tension Driven Flows,''
SIAM J. Appl. Math. {\bf 43}, 268 (1983).

\bibitem{/TK/} L. Ting and J. B. Keller, ``Slender jets and thin sheets with 
surface tension'', SIAM J. Appl. Math. {\bf 50}, 1533, (1990).

\bibitem{/BFL/} S. E. Bechtel, M. G. Forest, and K. J. Lin,
``Closure to all orders in 1-D models for slender viscoelastic free
jets: An integrated theory for axisymmetric, torsionless flows,''
SAACM {\bf 2}, 59 (1992).
  
\bibitem{/L/} H. C. Lee, ``Drop formation in a liquid jet'',
IBM J. Res. Develop.  {\bf 18}, 364 (1974).

\bibitem{/P95/} D. T. Papageorgiou, 
``Analytical description of the breakup of liquid jets'',
J. Fluid Mech. {\bf 301}, 109 (1995).

\bibitem{/FV/} M. A. Fontelos and J. J. L. Vel\'azquez,
``On the breakup of thin fluid tubes'', submitted.

\bibitem{/S3/} R. M. S. M. Schulkes, ``Nonlinear dynamics of 
liquid columns: A comparative study'',
Phys. Fluids. A {\bf 5}, 2121 (1993).

\bibitem{/M/} J. Meseguer, ``The breaking of axisymmetric
slender liquid bridges'', J. Fluid Mech. {\bf 130},
123, (1983).

\bibitem{/MS/} J. Meseguer and A. Sanz, ``Numerical and
experimental study of the dynamics of axisymmetric slender
liquid bridges'', J. Fluid Mech. {\bf 153},
83, (1985).

\bibitem{/S2/} R. M. S. M. Schulkes, ``Dynamics of liquid 
jets revisited'',
J. Fluid Mech. {\bf 250}, 635 (1992).

\bibitem{/PO/} D. T. Papageorgiou and O. Orellana,
``Study of cylindrical jet breakup using one-dimensional
approximations of the Euler equation'',
preprint.

\bibitem{/BO/} C. M. Bender and S. A. Orszag, {\em Advanced
mathematical methods for scientists and engineers}, 
(Mc Graw-Hill, N. Y., 1978).

\bibitem{/R/} I am grateful to an anonymous referee for this 
remark.

\bibitem{/UE/} Ch. Uhlig and J. Eggers, 
``Singularities in cascade models of the Euler equation'',
Z. Phys. B {\bf 103}, 69 (1997).

\bibitem{/CLS/} R. E. Caflisch, X. Li, and M. J. Shelley,
``The collapse of an axi-symmetric, swirling vortex sheet'',
Nonlinearity {\bf 6}, 843 (1993).

\bibitem{/PS/} `M. C. Pugh and M. J. Shelley, 
``Singularity Formation in Models of Thin Jets with Surface Tension'', 
Comm. Pure Appl. Math. {\bf 51}, 733 (1998).


\bibitem{/LT/} J. Lowengrub and L. Truskinovski
``Quasi-incompressible Cahn-Hilliard Fluids and 
Topological Transitions'', Proc. Roy. Soc. London, submitted

\bibitem{/Kr/} R. Krasny, ``A study of singularity formation in a 
vortex sheet by the point-vertex approximation'',
J. Fluid Mech. {\bf 167}, 65 (1986).



\end{references}
\end{document}